\documentclass[11pt]{article}
\usepackage{theorem}
\usepackage{latexsym}
\usepackage{amsmath}

\usepackage[dvips]{graphicx}
\usepackage{graphicx}

\theoremheaderfont{\scshape}
\newtheorem{theorem}{Theorem}[section]
\newtheorem{corollary}{Corollary}[section]
\newtheorem{lemma}{Lemma}[section]
\newtheorem{proposition}{Proposition}[section]
\newtheorem{postulate}{Postulate}[section]
\newtheorem{statement}{Statement}[section]
{\theorembodyfont{\rmfamily}
\newtheorem{example}{Example}[section]}
{\theorembodyfont{\rmfamily}
\newtheorem{remark}{Remark}[section]}
{\theorembodyfont{\rmfamily}
\newtheorem{definition}{Definition}[section]}

\newcommand{\newc}{\newcommand}

\newc{\be}{\begin{equation}}
\newc{\ee}{\end{equation}}
\newc{\bea}{\begin{eqnarray}}
\newc{\eea}{\end{eqnarray}}
\newc{\beas}{\begin{eqnarray*}}
\newc{\eeas}{\end{eqnarray*}}

\newc{\pardt}{\partial_{t}}
\newc{\pardxi}{\partial_{i}}
\newc{\pardts}{\partial_{t^{*}}}
\newc{\pardxis}{\partial_{i^{*}}}
\newc{\pardxj}{\partial_{j}}
\newc{\pardxk}{\partial_{k}}
\newc{\pard}{\partial}

\newc{\s }{\overline}

\newc{\sect}{\section}
\newc{\subs}{\subsection}

\newc{\defi}{\definition}
\newc{\prop}{\proposition}
\newc{\rem}{\remark}
\newc{\lem}{\lemma}
\newc{\exa}{\example}
\newc{\theo}{\theorem}
\newc{\coro}{\corollary}
\newc{\post}{\postulate}
\newc{\state}{\statement}

\begin{document}
\baselineskip0.5cm
\renewcommand {\theequation}{\thesection.\arabic{equation}}
\title{Alternative Vinen's equation and its extension\\ to rotating counterflow
 superfluid turbulence}

\author{M. Sciacca$^1$,
M.S.~Mongiov\`{\i}$^1$\thanks{Corresponding author.}  and
D.~Jou$^2$}

\date{ }
\maketitle
\begin{center} {\footnotesize $^1$ Dipartimento di Metodi e Modelli Matematici
Universit\`a di Palermo, c/o Facolt\`{a} di Ingegneria,\\ Viale delle Scienze, 90128 Palermo, Italy\\
$^2$ Departament de F\'{\i}sica, Universitat Aut\`{o}noma de
Barcelona, 08193 Bellaterra, Catalonia, Spain}

PACS number(s): 67.40.Vs, 67.40.Bz, 47.27.2i, 05.70.Ln
\end{center} \footnotetext{E-mail addresses: msciacca@unipa.it
(M. Sciacca), mongiovi@unipa.it (M. S. Mongiov\`{\i}),
david.jou@uab.es (D. Jou)}

\begin{abstract}
Two alternative Vinen's evolution equations for the vortex line
density $L$ in counterflow superfluid turbulence, are physically
admissible and lead to analogous results in steady states. In
Phys. Rev. B, 69, 094513 (2004) the most used of them was
generalized to counterflow superfluid turbulence in rotating
containers. Here, the analogous generalization for the alternative
Vinen's equation is proposed. Both generalized Vinen's equations
are compared with the experimental results,  not only in
steady-states but also in some unsteady situations. From this
analysis follows that the solutions of the alternative Vinen's
equation tend significantly faster to the corresponding final
steady state values than the solutions of the usual Vinen's
equation, and that the latter seems more suitable for the
description of the experimental available data.

\end{abstract}

\section{Introduction}
There is experimental evidence that turbulent helium II has a
peculiar behavior due to quantum effects at some length scales,
while at other length scales it appears similar to classical
hydrodynamic turbulence \cite{VN}-\cite{D1}. Quantum turbulence is
described as a chaotic motion of quantized vortices in a
disordered tangle. The measurements of vortex lines are described
in terms of a macroscopic average of the vortex line length per
unit volume $L$ (briefly called {\it vortex line density} and
which has dimensions $length^{-2}$).

 The evolution equation for $L$ under constant values of the counterflow velocity $\bf V$
 ($\bf V=<{\bf v}_n-{\bf v}_s >$, ${\bf v}_n$ and ${\bf v}_s$ being
the velocities of the normal and superfluid components) and in
absence of rotation was formulated by Vinen. Neglecting the
influence of the walls, such an equation is \cite{V2}:

\be\label{Vin} \frac{dL}{dt}  = \alpha V L^{3/2}   - \beta \kappa
L^2, \ee
 with $V=|\bf V|$, $\kappa=h/m$ the quantum of
vorticity ($m$ the mass of the $^4$He atom and $h$ Planck's
constant, $\kappa \simeq 9.97$ 10$^{-4}$cm$^2$/s) and $\alpha$ and
$\beta$ dimensionless parameters.

 Note however that another version of (\ref{Vin}) is the so-called
 alternative Vinen's equation, which is also admissible on
dimensional grounds \cite{V3}-\cite{NSF}:

\be \label{Vin-al} {dL\over dt} =A_1 {V^2\over\kappa} L-\beta \kappa
L^2 . \ee

The steady state solutions of (\ref{Vin}) and (\ref{Vin-al}) are
$L=(\alpha V/\beta\kappa)^2$ and $L=A_1V^2/\beta \kappa^2$
respectively, in agreement with the experimental results in
completely developed turbulent regime, which lead to
$L^{1/2}=\gamma V / \kappa$, with $\gamma$ a dimensionless
coefficient which depends on the temperature. Therefore, the
difference between (\ref{Vin}) and (\ref{Vin-al}) must be searched
in the dynamical aspects. This was carried out by Vinen himself
(see sections 6 and 7 of Ref.~\cite{V3}) and in more detail by
Nemirovskii {\it et al.} \cite{NSF} without arriving to definite
conclusions, because the predictions of (\ref{Vin}) and
(\ref{Vin-al}) in the domain of available experimental results are
very similar to each other.

Here we will look for a more general situation where the
difference between (\ref{Vin}) and (\ref{Vin-al}) becomes
enhanced. Essentially, equations (\ref{Vin}) and (\ref{Vin-al})
may be obtained from a microscopic approach based on vortex
dynamics, though the derivation of the first one is more direct
and straightforward than the other. Before proceeding, let us
briefly recall that from a microscopic approach based on vortex
dynamics the production term may be shown to be proportional to
$VL/R_{peak}$, where $R_{peak}$ is the intervortex spacing
\cite{NF,Schwartz}. Since the latter is of the form $L^{-1/2}$,
the form $VL^{3/2}$ adopted in (\ref{Vin}) follows in an immediate
way. However, it has been also argued that \cite{NF}, since in the
steady state $L^{1/2}$ is proportional to $V$, one could also
write $R_{peak}$ as inversely proportional to $V$, in which case
one would have for the production term the alternative form $V^2L$
adopted in (\ref{Vin-al}). The equation (\ref{Vin-al}) was also
derived by Lipniacki (pag. 177 of Refs.~\cite{BDV,Li}) through an
alternative microscopic approach, which is referred to the
reconnections of vortex lines.

Let us say, in support of the form $V^2L$, that though it is less
natural from a kinematical point of view, it is rather natural
from an energetic point of view, if one assumes in agreement with
the microscopical model$^{1,2}$ that the friction force between
the normal fluid and the vortex tangle is proportional to $VL$. In
this case, the power delivered to the tangle per unit volume would
be proportional to $V^2L$. Since the energy density of the tangle
is proportional to $L$, the production contribution to $dL/dt$
would be proportional to the power delivered to the tangle, i.e.
to $V^2L$.

Another motivation in support of (\ref{Vin-al}) is that it is
closer to the phenomenological theory of classical turbulence
\cite{LL} than equation (\ref{Vin}). Indeed, by assuming that
turbulence can be characterized by the line density $L$ and that
its derivative $dL/dt$ is an analytic function of $L$, the
relation $A_1V^2L$ can be interpreted as the first term in a
series expansion. However, equation (\ref{Vin}) has been much more
used than equation (\ref{Vin-al}). Both (\ref{Vin}) and
(\ref{Vin-al}) are particular cases of:

\be \label{Vin2} \frac{d L}{dt}=A_n\kappa L^2\left(\frac{V}{\kappa
L^{1/2}}\right)^n-\beta \kappa L^2, \ee in fact for $n=1$ one
obtains (\ref{Vin}), for $n=2$ one has (1.2), and the cases $1\leq
n \leq 2$ could correspond to fractal (intermittent) turbulence
\cite{JLM}-\cite{JM-libro}.

In recent years there has been growing attention in superfluid
turbulence in rotating containers \cite{F1}-\cite{TBAM}, in which
the formation of vortex lines is due both to the counterflow and
the rotation, which has fostered the extension of Vinen's ideas to
a wider range of situations \cite{JM-libro,JM1,JM2}. In
Ref.~\cite{JM1}, a phenomenological generalization of Vinen's
equation (1.1) has been proposed for the evolution of $L$ in the
simultaneous presence of $\bf V$  and $\bf\Omega$ ($\bf\Omega$
being the angular velocity of the container) . A thermodynamic
analysis to determine possible coupling terms between the
evolution equations of $L$ and $\bf V$ has been performed in
Refs.~\cite{JM-libro}.

Here, the extension of the form (\ref{Vin-al}) of Vinen's equation
to rotating counterflow turbulence will be studied in order to
explore whether this more general situation may provide further
arguments to decide which of both starting equations, (\ref{Vin}) or
(\ref{Vin-al}), is more suitable to describe actual experimental
results. Of course, the final version will be fully satisfactory
only when the macroscopic expression will be understood from a
microscopical basis, and the respective coefficients of all terms
will be microscopically calculated and found to coincide with
macroscopic observations. However, this situation is still far ahead
of our present abilities, because of the difficulties in modeling
--- in a statistically significative way --- a system of vortices
under rotation. Thus, a combined effort in macroscopic and
microscopic perspectives seems a reasonable and promising way to
proceed.

The plan of the paper is the following. In Section 2 a new
equation for the evolution of $L$ in counterflow in rotating
containers is written, through a modification of the Vinen's
alternative equation, and it is solved in steady and unsteady
situations in order to compare it with the generalization of the
usual Vinen's one made in Ref.~\cite{JM1}. In Section 3 a
thermodynamic analysis of counterflow rotational superfluid
turbulence is performed, according to the formalism of
nonequilibrium thermodynamics, to obtain the general form of the
friction exerted by the tangle on the motion of the fluid.

\section{New equation for the dynamics of
$L$ in rotating counterflow superfluid turbulence}
\setcounter{equation}{0}

There are not many experiments on counterflow in rotating
containers. In the work of Swanson {\it et al.} \cite{SBD}, the
counterflow velocity $V$ was parallel to the rotation axis and the
experimental observations consisted in measuring the attenuation
of second sound, when it is propagated orthogonal to the rotation
axis. They interpreted their results as measurements of the vortex
line density $L$, and compared the observed line density with what
would be expected if the  two sources  of vorticity (rotation and
counterflow) simply added. Their results showed an interesting
interplay between the ordered vortices of rotation and the
disordered ones of counterflow. They found a critical counterflow
velocity which marks the transition to a disordered turbulent
tangle. In the limit of high rotation this critical velocity
appears proportional to $\sqrt \Omega$.

Another experiment is that carried out by Yarmchuck and Glaberson
\cite{YG}, in which a pair of horizontal parallel glass plates are
arranged to form a large aspect ratio channel closed at one end
with a heater nearby, and open at the other end to the liquid
helium bath.  The channel is rotated about a vertical axis
orthogonal to the heat flux. In this way the counterflow velocity
$\bf V$ is orthogonal to angular velocity $\bf\Omega$. By
determining gradient of temperature and chemical potential as
functions of $V$ and $\Omega$, they found a linear regime, in
which these gradients  grow proportionally to the rotation speed,
and a critical counterflow velocity associated with the onset of
turbulent regime, which increases as $\sqrt\Omega$ when $\Omega$
gets large.

\subsection{The evolution equation}\label{sez311}

The mentioned experiments by Swanson, Barenghi and Donnelly
\cite{SBD} and by Yarmchuck and Glaberson \cite{YG} show that,
when the heat flux and the rotation are simultaneously present,
there appears a complex interaction between both processes in the
formation and destruction of vortices. In the experiment performed
in Ref.~\cite{SBD} ($\bf\Omega$ and $\bf V$ collinear), they
observed that the effects of $V$ and $\Omega$ are not additive: in
fact, for low values of $\Omega$, the laminar regime (vortex-free
regime) is absent and the total vortex line density is higher than
$L_R+L_H$, $L_R$ and $L_H$ being the values of $L$ in steady
rotation and in steady counterflow superfluid turbulence
respectively:

\be\label{LRLH} L=L_R={2\Omega \over\kappa},\hskip0.5in L_H=
\gamma^2 \frac{V^2}{\kappa^2}, \ee with $\gamma$ a dimensionless
coefficient while, for higher values of $\Omega$ and $V$, the
measured values of $L$ are always less than $L_H+L_R$, and the
deviation increases with $V$ and $\Omega$. Therefore, the rotation
facilitates the vortex formation, in the absence or for small
counterflow velocities, but it hinders their lengthening for high
values of $V$ and  $\Omega$.

For fast enough rotation, there are two critical
counterflow-rotation velocities $V_{c1}$ and $V_{c2}$, which scale
as $\Omega^{1/2}$ ($V_{c1}= C_1\sqrt{\Omega}$, $V_{c2}=
 C_2\sqrt{\Omega}$, with $C_1=0.053$ cm~sec$^{-1/2}$, $C_2=0.118$
 cm~sec$^{-1/2}$).
For $V \le V_{c1}$, the length $L$ per unit volume of the vortex
lines is independent of $V$ and agrees with the first expression in
(\ref{LRLH}). For $V_{c1} \le V \le V_{c2}$, $L$ is still
independent of $V$ and proportional to $\Omega$, with a slightly
different proportionality constant than in the previous situation;
finally, for $V \ge V_{c2}$, $L$ increases and becomes proportional
to $V^2$ at high values  of $V$.

Swanson {\it et al.} \cite{SBD} interpreted the first transition
as the Donnelly-Glaberson instability {CCD,OG}: excitation of
helical waves (Kelvin waves) by the counterflow on the vortex
lines induced by rotation, and the second as a transition to a
turbulent disordered tangle. Tsubota {\it et al.} \cite{TBAM} also
have paid attention to this experiment. They proposed that the
regime $V_{c1}<V<V_{c2}$ is a state of polarized turbulence, while
for $V>V_{c2}$ the polarization is decreased by the large number
of reconnections.

In the regime of high rotation ($0.2$ Hz $\le \Omega /2\pi \le
1.0$ Hz and $0\le V^2\le 0.2$ cm$^2$/s$^2$) and $\bf \Omega$
parallel to $\bf V$, equation (\ref{Vin}) has been generalized to
incorporate the presence of rotation, as \cite{JM1}:

\be\label{VinEqua22} {dL\over dt}= -\beta  \kappa  L^2 +
\alpha_1\left[L^{1/2} - m_1 \frac{
\sqrt{\Omega}}{\sqrt{\kappa}}\right]V L + \beta_2\left[L^{1/2} -
m_2 \frac{ \sqrt{\Omega}}{\sqrt{\kappa}}\right]\sqrt{\kappa\Omega}
L, \ee where $m_1$ and $m_2$ are linked to the coefficients
introduced in  Ref.~\cite{JM1} by the relations $m_1= {\beta_4}
/{\alpha_1}$ and $m_2= \beta_1/\beta_2$, with the coefficients
$\beta$, $\alpha_1$, $\beta_1$, $\beta_2$ and $\beta_4$ depending
on the polarization of the tangle, which was supposed function of
$\Omega$ and $V$.

Note that, as production terms in Eq. (\ref{VinEqua22}) a term in
$V$ and a term in $\Omega^{1/2}$ were used; this was motivated by
the dependence of the steady-state values of $L^{1/2}$, in
counterflow only and in rotation only, on $V$ and on
$\Omega^{1/2}$ (see equations (\ref{LRLH})), and by the
observation that   the microscopic mechanism responsible for the
growth of vortices (the mutual friction force) is the same in
rotating helium II  and  in superfluid turbulence. There are
present three destruction contributions: a term $-\beta\kappa L^2$
independent on $ V$ and on $\Omega$, present also in (\ref{Vin})
and (\ref{Vin-al}) (this term, responsible for the  vortex  decay
in pure counterflow, was determined  by Vinen in analogy with
classical turbulence)  and the two terms $- \beta_1 \Omega L$ and
$-\beta_4 \frac{V \sqrt{\Omega}}{\sqrt{\kappa}}L$, which take
account of the interactions between counterflow and rotation,
reducing the length of the vortices; a term quadratic in $V$ was
neglected, because the values of $V$ used in the experiments by
Swanson {\it et al.} were not very high.

The Eq. (\ref{VinEqua22}) describes, in good agreement with
experimental results, some of the most relevant effects observed
in the experiments of Ref.~\cite{SBD}. However, as we have
mentioned in the Introduction, the alternative Vinen's equation
also describes well the experimental results in pure counterflow
and therefore it is natural to ask how does it work when extended
to incorporate rotation.

Here, we suggest a new evolution equation for the evolution of
vortex line density $L$ in rotating counterflow, starting from the
alternative Vinen's equation and following the lines of thought
outlined in Ref.~\cite{JM1}. We consider the case in which $V$ and
$\Omega$ are parallel to each other. The proposed new equation,
reducing to (1.2) for vanishing rotation, is:
 \be\label{AltVinEqua} {dL\over dt} = -\beta  \kappa
L^2 +A_1 \left[L - \nu_1 {\Omega\over {\kappa}}\right]{V^2 \over
\kappa}
 +  B_1 \left[ L- \nu_2{\Omega\over {\kappa}}\right] {\Omega},
  \ee
where the coefficients $\beta$, $A_1$, $\nu_1$, $B_1$ and $\nu_2$
depend on the polarization of the tangle, which was supposed
function of $\Omega$ and $V$.

 We outline now a possible physical interpretation for the
terms of production and destruction of vortices introduced in this
equation. The two production terms, $A_1 {V^2 \over \kappa}L$ and
$B_1  {\Omega}L$, indicate that both rotation and counterflow
favor the vortex formation:  the quantities $t_H=\kappa/( A_1
V^2)$ and $t_R=1/(B_1 {\Omega})$ can be interpreted as the
characteristic times for the formation of vortex lines due to the
counterflow and to the rotation, respectively. As in
(\ref{VinEqua22}), three destruction terms are present. The term
$A_1 \nu_1 {\Omega }{V^2 /\kappa^2}$ describes the complex
interaction between rotation --- which tends to straighten out the
otherwise irregular vortex lines of the tangle, thus shortening
them and reducing $L$ --- and counterflow, which randomize them.
Another aspect especially worth of comment is the meaning of the
two destruction terms, independent on $V$, $-B_1\nu_2 \Omega^2
/\kappa$ and $-\beta\kappa L^2$. One could argue, indeed, that at
steady pure rotation there is no vortex destruction. Thus, in
purely rotation situations the vortices are usually produced on
the walls and they migrate to the bulk of the fluid in the
cylinder; in this case, these terms  would represent a repulsion
force between parallel vortices, putting an upper limit to the
possible number of straight vortices in the vortex array.

In a general situation, these two destruction terms will
incorporate real destruction of vortices due to breaking
recombination of nonparallel vortices, and to repulsion between
parallel segments of vortices in the presence of rotation. The
superposition of these two different effects is one of the reasons
that the coefficients in the terms in $L^2$ and in $\Omega^2$
depend on the polarization of the tangle. The destruction term
$-\beta\kappa L^2$, which appears also in (\ref{Vin}),
(\ref{Vin-al}) and (\ref{VinEqua22}), is not modified, in
agreement with recent studies which show that the decay of the
turbulence, in the absence of rotation and counterflow velocity,
is analogous to that of classical turbulence \cite{LL}.

As it was shown in Ref.~\cite{JM1}, the simplicity of
(\ref{VinEqua22}) as (\ref{AltVinEqua}) is a little bit deceptive,
because the coefficients appearing in it depend on the polarity
and the anisotropy of the vortex tangle, which are taken as
independent variables in the more detailed approach proposed in
the recent paper \cite{JLM}. Of course, this difficulty arises not
only in the macroscopic approach, but also in the microscopic
approaches. Since the coefficients appearing in (\ref{AltVinEqua})
depend on the anisotropy and on the polarization of the tangle,
they have to depend on the angular velocity $\Omega$ and on the
counterflow velocity $V$. In particular, when $\Omega=0$ Eq.
(\ref{AltVinEqua}) reduces to (1.2), and the coefficient
$A_1/\beta$ assumes the value: \be\label{A1}
\frac{A_1}{\beta}=0.156.\ee In the analysis carried out in the
present paper, the dependence of the coefficients on the polarity
plays not an important role, because in Section~2.2 we are
comparing two equations in a regime of values of $V$ and $\Omega$
($0.2$ Hz $\le \Omega /2\pi \le 1.0$ Hz and $0\le V^2\le 0.2$
cm$^2$/s$^2$) with approximately the same polarity, and in
Section~2.4 we compare the non-stationary behavior of the
perturbations to a given physical situation, as described by two
different equations.

In the successive subsections we solve equation (\ref{AltVinEqua})
in steady and unsteady situations and we show that it allows us to
account for the experiments described in Ref.~\cite{SBD} and we
compare the results with the description pointed in
Ref.~\cite{JM1} based on the extension of (1.1).

\subsection{The stationary solutions and their stability}

The non zero stationary solutions of (\ref{AltVinEqua}) are
solutions  of the following second-order algebraic equation in the
unknown $L$:

\be\label{algequa} \beta  \kappa  L^2 - \left[ {A_1 \over \kappa}
V^2 +B_1 {\Omega} \right] L + \left[ B_2{\Omega^2\over
{\kappa}}+{B_3\over \kappa^2} V^2\Omega \right] =0, \ee where we
have put $B_2= B_1\nu_2$ and $B_3=A_1\nu_1$.

Looking at the experimental results of Ref.~\cite{SBD}, one notes
that $L$ is almost independent of $V$ for $V<V_{c2}$, with a step
change around $V_{c1}$, while there is a variation of the slope
near $V_{c2}$. We will concentrate on the change near $V_{c2}$.
Reasoning as in Ref.~\cite{JM1}, we observe that, under the
hypothesis:

\be {B_2\over\beta}={B_3\over A_1}\left({B_1\over\beta}- {B_3\over
A_1} \right), \ee it follows that:

\be {B_1\over\beta} = {\nu_1^2 \over \nu_1-\nu_2} ,\hskip0.4in
{B_2\over\beta}= {\nu_1^2 \nu_2\over \nu_1-\nu_2},
 \ee and the solutions of equation (\ref{algequa}) can be written:

\be\label{solstazL1} L=L_1^A=\nu_1{\Omega\over\kappa}, \ee
\be\label{solstazL2} L=L_2^A= {A_1\over\beta}
{V^2\over\kappa^2}+\left({B_1\over \beta}
-\nu_1\right){\Omega\over\kappa} .
 \ee

\begin{figure}[h]
   \includegraphics[width=10cm]{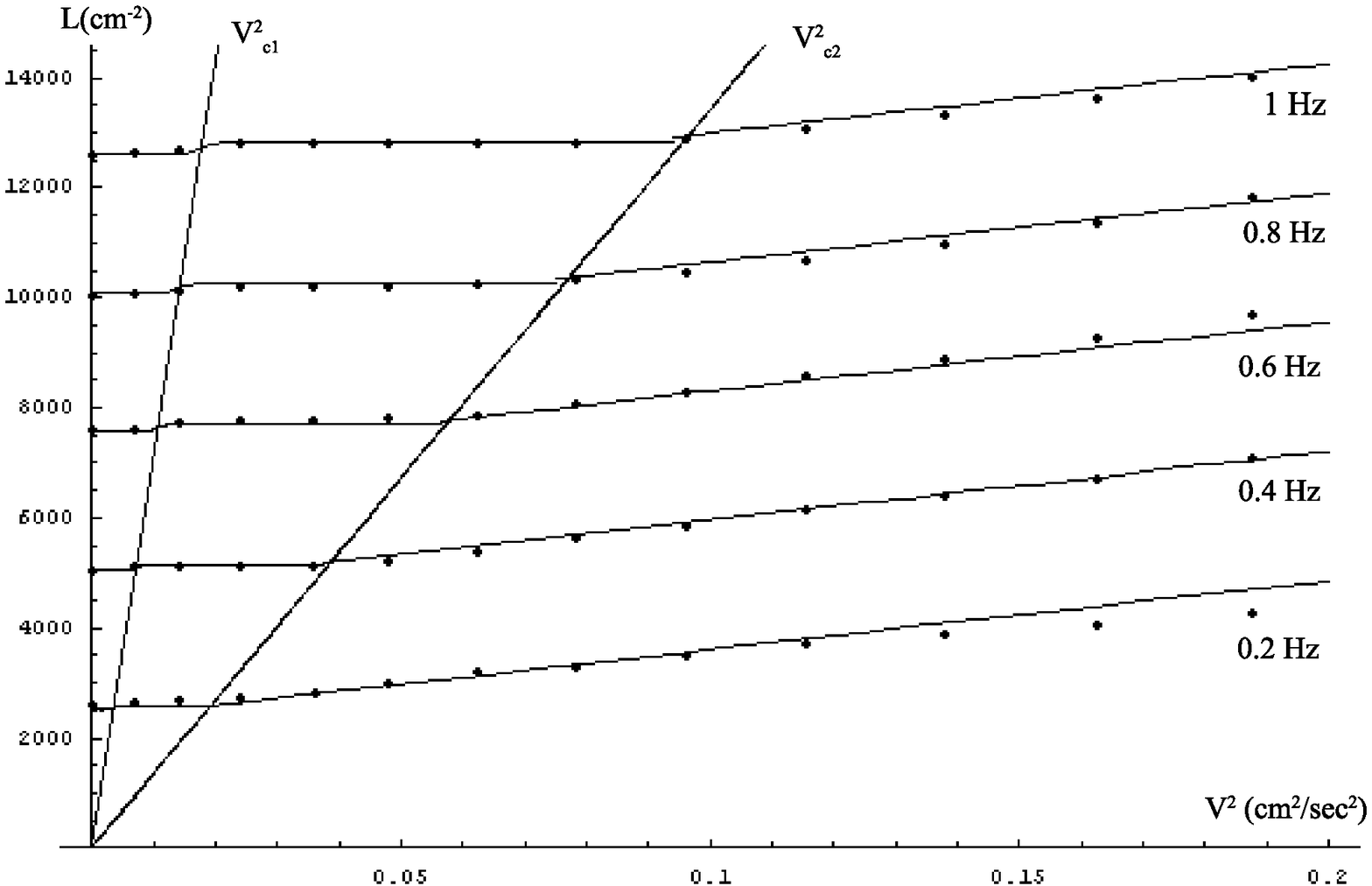}\\
  \caption{Comparison of the solutions (\ref{solstazL1AValue})
  (continuous line) with the experimental data by Swanson {\it et al.} \cite{SBD}.}
  \label{confronto}
\end{figure}

 In the plane $(V^2,L)$, (\ref{solstazL1}) and (\ref{solstazL2}) represent two families of
straight lines plotted in Fig.~\ref{confronto}, the first of them
(equation (\ref{solstazL1})) parallel to the $V^2$ axis and the
second one (equation (\ref{solstazL2})) with a slope independent of
$\Omega$. A linear stability analysis of these solutions shows that
the solution (\ref{solstazL1}) is stable if $V$ is lower than:

\be\label{VeloCrit2} V_{c2}^2={\beta\over A_1} \left[2{B_3\over
A_1}-{B_1 \over \beta}\right] {\Omega \kappa} = {\beta\over A_1}
{\nu_1^2-2\nu_1\nu_2 \over \nu_1-\nu_2} {\Omega \kappa}, \ee
(corresponding to the point of interception of  the  two  straight
lines (\ref{solstazL1}) and (\ref{solstazL2})), while, for values
of $V$ higher than $V_{c2}$, the solution (\ref{solstazL2}) is
stable. Therefore $V_{c2}$ represents the second critical
counterflow-rotation velocity observed in the experiments of
Ref.~\cite{SBD}.
  As we see, this critical velocity
scales as $\sqrt{\Omega}$, in agreement with experimental
observations.

The experimental data on the steady states of $L$ allow us to
determine the  values assumed by the dimensionless quantities
appearing in equation (\ref{AltVinEqua}). One obtains:

\be\label{ValueCosta} {A_1\over\beta}=0.0125 , \hskip0.2in {B_1
\over\beta}= 3.90, \hskip0.25in {B_2 \over\beta}=3.79  , \hskip0.2in
{B_3 \over\beta}= 0.025, \ee from which we obtain:
\be\label{Valuealtri} \nu_1=2.036, \hskip0.4in \nu_2=0.97. \ee

  The coefficient $\beta$, which controls the rate of evolution of $L$,
cannot be determined from the knowledge of the stationary solutions.
Comparing the value $A_1/\beta$ obtained in the combined situation
with the value (\ref{A1}) obtained in absence of rotation, we can
deduce that the first one is approximately 12 times the second one,
which means that the coefficient $A_1/\beta$ depends on the
anisotropy and the polarity of the tangle, which depend on $\Omega$
and $V$.

Using the obtained values of the dimensionless quantities
(\ref{ValueCosta}) and (\ref{Valuealtri}), the steady stationary
solutions $L_1^A$ and $L_2^A$ become:

\be\label{solstazL1AValue} L_1^A=2.036{\Omega\over\kappa},
\hspace{1cm} \textrm{and} \hspace{1cm} L_2^A= 0.0125
{V^2\over\kappa^2}+1.86{\Omega\over\kappa}.
 \ee
In Fig.~\ref{confronto} a comparison of such stationary solutions
$L_1^A$ and $L_2^A$ with the experimental data of Swanson {\it et
al.} is shown \cite{SBD}. The conclusion of such a fit is that the
stationary vortex line density $L_1^A$ and $L_2^A$, solutions of
the alternative Vinen's equation in the combined situation, are in
good agreement with experimental data of Swanson {\it et al.}

In Ref.~\cite{JM1}, the stationary solutions of equation
(\ref{VinEqua22}) had the form:

\be\label{solstazL1Value}
L_1^{1/2}=1.427\sqrt{\frac{\Omega}{\kappa}} \hspace{1cm}
\textrm{and} \hspace{1cm} L_2^{1/2}=0.047
\frac{V}{\kappa}+1.25\sqrt{\frac{\Omega}{\kappa}}  \ee and the
comparison with the experimental data led also to the conclusion
that (\ref{solstazL1Value}) agree with the experiments by Swanson
{\it et al.} \cite{SBD}. Through (\ref{solstazL1AValue}) and
(\ref{solstazL1Value}) have a different mathematical form, in the
range of the available experimental data, both of them lead to
reasonable results.

From such conclusions an interesting problem is to establish which
equation, either (\ref{VinEqua22}) based on the usual Vinen's
equation or (\ref{AltVinEqua}) based on the alternative Vinen's
equation, fits better the experimental data obtained by Swanson,
Donnelly and Barenghi \cite{SBD}.

From a first comparison, the two stationary solutions
(\ref{solstazL1Value}a) and (\ref{solstazL1AValue}a) represent the
same straight line in the plane $(L,V^2)$ in the range
$V^2_{c1}<V^2<V^2_{c2}$. So, an eventual difference between both
equations could be found in the range $V^2>V^2_{c2}$. To do that, we
calculate the errors $\sigma$  between $L_2$ and the corresponding
experimental value $\overline{L}$, and $\sigma^A$ between $L_2^A$
and $\overline{L}$, respectively, in such a way that we can compare
the accuracy of the two models.

To find these errors, we consider the experimental values $V_i^2$
and $\Omega_j$ of the experiments to which $L_{2ij}$, $L_{2ij}^A$
and $\overline{L}_{ij}$ correspond, obtaining:

\be\label{error1} \sigma=\sqrt{\frac{\sum_{i,j}
\left(L_{2ij}-\overline{L}_{ij}\right)^2}{N}}=963 \ee and
\be\label{error2} \sigma^A=\sqrt{\frac{\sum_{i,j}
\left(L_{2ij}^A-\overline{L}_{ij}\right)^2}{N}}=419, \ee where $N$
is the number of experimental data, which is equal for both cases.
From (\ref{error1}) and (\ref{error2}) we can establish that the
stationary solution of the alternative Vinen's equation approaches
better the experimental data (for $V^2>V_{c2}^2$) than that of the
usual Vinen's equation.

\subsection{The first critical velocity}

The model based on the equation (\ref{AltVinEqua}) does not describe
the existence of the first critical velocity $V_{c1}$ mentioned in
Section (\ref{sez311}), in which the value of $L$ has a small steep
change. To do this, we assume that the coefficient $\nu_1$ depends
on $\Omega$ and $V$ as: \be\label{espreni1}\nu_1= A\left\{1-B \tanh
\left[ N'\left({ {k\Omega} \over V^2} -C\right)\right] \right\}, \ee
 with $A$, $B$ and $C$ constants. Thus,
for $V^2 \ll V_{c1}^2  =  {1\over C}{k\Omega}$, it results
$\nu_1\simeq A-BA$ and for $V^2\gg V_{c1}^2$,  $\nu_1=A+BA$, while
the constant $C$ is related to $V_{c1}$, and $2B$ gives the size of
the step of $\nu_1$ near $V_{c1}$. In fact, if $V_{c1}$ is small,
the domain of $V$ in which  the mentioned transition occurs is very
narrow, as observed in experiments.

In (\ref{espreni1}), the critical value  $V_{c1}$  of  the
counterflow velocity is given in terms of coefficient $C$ by: \be
V_{c1}^{2}={1\over C}  {\kappa\Omega}. \ee  Using the experimental
values   of   $V_{c1}$ ($V_{c1} =0.053 \sqrt\Omega$
cm~sec$^{-1/2}$), it is seen that $C=\kappa /(0.053)^2 =0.355$.

To determine the coefficients  $A$  and  $B$ in (\ref{espreni1}),
we consider that for a given value of $\Omega$, for small values
of $V$, the tangle will be completely oriented along the rotation
axis, and $\nu_1=A-BA=2$. On the other side,  when $V\gg V_{c1}$
(i.e. near $V_{c2}$) $\nu_1$ assumes its higher value
$\nu_1^{max}$ furnished by $A+BA=\nu_1^{max}$; the value of
$\nu_1^{max}$ was obtained in Ref.~\cite{JM1} using experimental
data of Ref.~\cite{SBD}, and is $\nu_1^{max}=2 L(V_{c2})/L_R =
2.036$. It is seen that the step in $\nu_1$ is indeed small. In
(\ref{espreni1}), $N'$ is a phenomenological coefficient
characterizing the rate of growth of $L$  near $V_{c1}$ and the
experimental data show that $N'> 20$, but do not allow to
determine it with precision. Here, we will chose for it the value
$N'=22$ proposed by Tsubota {\it et al.} in Ref.~\cite{TBAM}.

Expression (\ref{espreni1}) is similar to that proposed in
Ref.~\cite{JM1} and it is founded on the microscopic ideas about
the nature of the transition, already proposed by Donnelly
\cite{D1}, according to which, for small $V$, the vortex lines are
straight lines parallel to the rotation axis, but increasing
values of $V$ produce helical perturbations of the vortex lines
around  their low-$V$ configuration. The situation has been
compared by Donnelly to magnetic systems, where the external field
$H$ contributes to the orientation  of magnetic dipoles, while the
temperature $T$ has a disordering effect. Thus, the "$\tanh$" term
in (\ref{espreni1}) is analogous to the expression describing
magnetization in terms of magnetic field and temperature in a
$1/2$ spin paramagnetic system. Other expressions, as for
instance, Langevin's one for classical paramagnetism, could also
be used \cite{TBAM,JM4,JM-PLA359}. This ansatz is similar to that
proposed in Ref.~\cite{JM5}, to explain the transition from the
laminar to the turbulent regime in pure counterflow, and is based
on an analogous physical basis: there, the flow was producing the
helicoidal excitation waves along the vortices pinned to the walls
of the container always present in the laminar regime. We recall
that in pure counterflow, in containers with circular and square
section, there are three different regimes distinguished by two
critical counterflow velocities: under the first critical velocity
we have the laminar regime where only a few of vortices pinned to
the walls of the containers are present, between the two critical
velocities a state of low vortex line density (TI regime) is
formed, and, at last, above the second critical velocity a state
of higher values of $L$ (TII regime) is present. In
Ref.~\cite{JM5} the transition from the laminar regime to the
turbulent TI regime was explained supposing that small localized
arrays of quantized vortices appear when the counterflow velocity
reaches the first critical velocity, because Kelvin waves may be
propagated in these pinned vortices. Whereas, when the counterflow
reaches the second critical velocity, the TI turbulent regime,
which is an inhomogeneous and locally polarized state, becomes
unstable, with a transition to an homogeneous slightly not
isotropic state TII.

In the microscopic  model we have commented on, the second
critical velocity $V_{c2}$ is interpreted as  the  velocity where
the helical vortex  lines  produced  in $V_{c1}$ have reached an
amplitude of the order of the average vortex separation and have
broken and reconnected, and form a disordered tangle. This
explanation is analogous to the one given in Ref.~\cite{JM5} to
explain the transition from TI to TII turbulent regimes.

\subsection{Non-stationary solutions of the Vinen's and alternative
Vinen's equations}

In this Subsection we study the non-stationary behavior of
(\ref{VinEqua22}) and (\ref{AltVinEqua}). Though the lack of
experiments about the evolution of the vortex line density $L$ in
this more general case (rotation and counterflow) does not allow
us to compare directly our results with experimental data, however
we can arrive at some interesting conclusions concerning the
difference of behavior.

First of all, we have to state that the analysis below refers to
$L$ as dependent variable when the growth of $V$ or of $\Omega$ is
very small. Two main situations are considered, in the first one
the angular velocity is fixed and the counterflow velocity moves
between two consecutive experimental values (see
Fig.~\ref{confronto}); in the second one the opposite situation is
assumed, that is $V$ is fixed and $\Omega$ grows in a small range.
This choice is due to the fact that when the ranges of $V$ and
$\Omega$ are sufficiently large, the coefficients of the Vinen's
equations may be not constant, as showed by Schwarz and Rozen in
Ref.~\cite{SR}, because they may depend on the anisotropy and
polarization of the tangle.

Denoting with $L_0$ the initial value of $L$, the solution of the
evolution equation (\ref{AltVinEqua}) is:

\be\label{SolNonStAlt} \beta \kappa t (L_2^A-L_1^A)=\ln \left|
\frac{(L-L_1^A)(L_0-L_2^A)}{(L_0-L_1^A)(L-L_2^A)}\right|, \ee with
$L_1^A$ and $L_2^A$ given by (\ref{solstazL1AValue}); while the
solution of
 the equation (\ref{VinEqua22}) can be written
as:

\begin{eqnarray}\label{SolNonSt}
 \nonumber  -\frac{\beta \kappa}{2} t =\frac{1}{\sqrt{L_1
L_2}} \ln \left|\frac{\sqrt{L}}{\sqrt{L_0}} \right|+
\frac{1}{\sqrt{L_1}(\sqrt{L_1}-\sqrt{L_2})} \ln
\left|\frac{\sqrt{L}-\sqrt{L_1}}{\sqrt{L_0}-\sqrt{L_1}} \right| \\
  + \frac{1}{\sqrt{L_2}(\sqrt{L_2}-\sqrt{L_1})} \ln
\left|\frac{\sqrt{L}-\sqrt{L_2}}{\sqrt{L_0}-\sqrt{L_2}} \right|,
\end{eqnarray}
where $L_0$ is the initial value of $L$, and $L_1$ and $L_2$ are
expressed by (\ref{solstazL1Value}).

In order to compare the unsteady solutions (\ref{SolNonStAlt}) and
(\ref{SolNonSt}) of (\ref{AltVinEqua}) and (\ref{VinEqua22}), a
value for the coefficient $\beta$ must be chosen. As already said,
 $\beta$ may depend on the anisotropy and
polarization of the tangle, therefore it may have a different
value with respect to the one in pure counterflow situation.
However, since this dependence is not known in this section, to
perform this comparison we choose the value of $\beta$ in pure
counterflow, namely $\beta=1/2\pi$.

\textbf{For  $\Omega$ fixed}.  Now, we choose some values for $V^2$
and $\Omega$ in order to plot the solutions of the two models. First
of all we consider the case $V^2<V_{c2}^2$, and in particular the
values $V^2=0.0072$ and $\Omega/2 \pi = 0.4$ to which  the following
values of the stationary solutions correspond:

\[L_1=L_1^A= 5132, \hspace{0.5cm}
L_2= 4455, \hspace{0.5cm} L_2^A= 4779.
\]
For the initial value $L_0$ we choose $L_0=L_R=2 \Omega/\kappa$.
Here, all the values for $L$, $V^2$ and $\Omega$ will be expressed
in $\textrm{cm}^{-2}$, $\textrm{cm}^{2}\ \textrm{s}^{-2}$ and
$\textrm{rad}\ \textrm{s}^{-1}$, respectively.

\begin{figure}[h]
   \includegraphics[width=10cm]{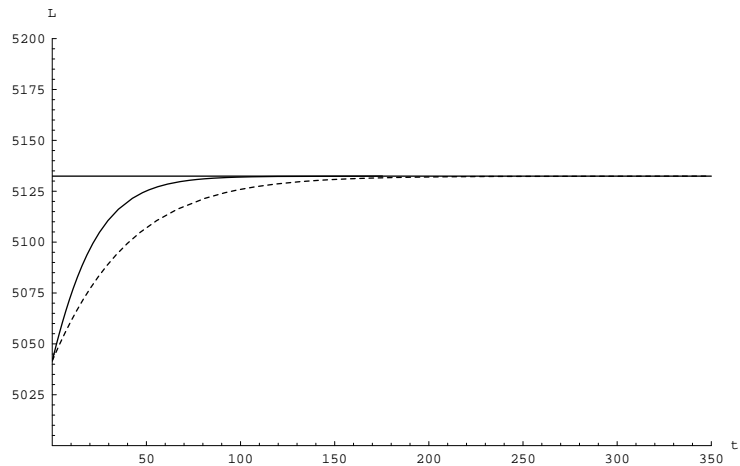}\\
  \caption{Evolution of the vortex line density $L$ towards its
  steady state value for the generalizations of the usual Vinen's
  equation (\ref{VinEqua22}) [dotted line] and the alternative
  Vinen's equation (\ref{AltVinEqua}) [continuous line] for $\Omega/2 \pi = 0.4$ and $V^2=0.0072$, lower than
  the critical value $V_{c2}$.}
  \label{lowerVc2}
 \end{figure}

From the analysis of Ref.~\cite{JM1} we already know that in this
range the stationary solution $L_1$ is stable. The same conclusion
is reached by looking at the plot of the non-stationary solutions
(\ref{SolNonStAlt}) and (\ref{SolNonSt}) of the two models in
Fig.~\ref{lowerVc2}. Further, we note that the values of $L_2$ and
$L_2^A$ are smaller than $L_1$ and that the non-stationary
solutions approach to the stable stationary one, $L_1$, in
relatively similar times.

\begin{figure}
   \includegraphics[width=10cm]{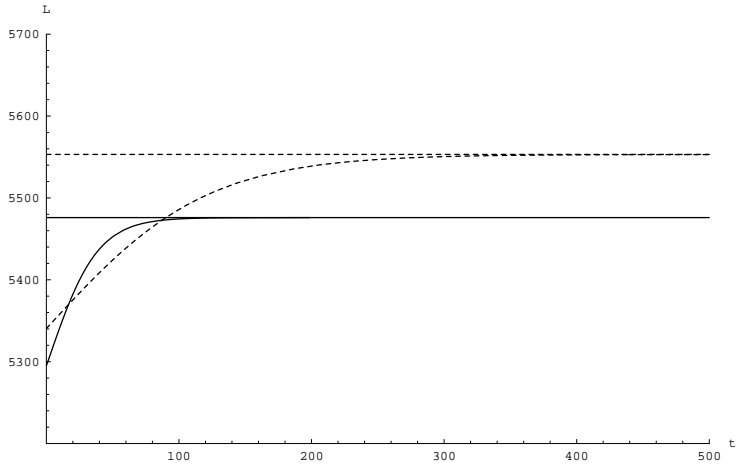}\\
  \caption{As in Fig.~\ref{lowerVc2}, but for $\Omega/2 \pi = 0.4$ and $V^2=0.0626$ slightly above the critical value $V_{c2}$.
  Note that the steady solutions differ only in a $0.75\%$, whereas the difference in the time necessary to reach the steady
  state differs in more than $300\%$. The values of $L_0$ are the unsteady solutions $L_2$ and  $L^A_2$ at
  the same $\Omega$ and $V^2=0.0482$.}
  \label{aboveVc212}
\end{figure}

Following the same process as above and setting the same value for
$\Omega$ and a value $V^2=0.0626$ slightly higher than $V_{c2}^2$,
we find the following values for the stationary solutions:

\[L_1=L_1^A= 5132, \hspace{0.5cm}
L_2= 5553, \hspace{0.5cm} L_2^A=5476,
\]
and for $L_0$  two different values $L_{2_{|V^2=0.0482}}$ and
$L_{2_{|V^2=0.0482}}^A$ are chosen respectively for the two
solutions (\ref{SolNonStAlt}) and (\ref{SolNonSt}) (see Fig.
\ref{aboveVc212})). Note that in this case the value of $L_1$ is
smaller than $L_2$ and $L_2^A$. As we know from previous studies,
in this range the stationary solutions $L_2$ and $L_2^A$ for the
Vinen's equation and alternative Vinen's one are stable. This is
confirmed in  Fig.~\ref{aboveVc212}, where (\ref{SolNonStAlt}) and
(\ref{SolNonSt}) are plotted.

In Fig.~\ref{aboveVc212} we also note a different behavior with
respect to that in Fig.~\ref{lowerVc2}; in fact, the two
non-stationary solutions $L$ approach the corresponding stationary
values $L_2^A$ and $L_2$ in rather different times with a ratio of
about 1:3, respectively. So, the solution of the alternative Vinen's
equation is faster than that of the Vinen's equation.

\begin{figure}
   \includegraphics[width=10cm]{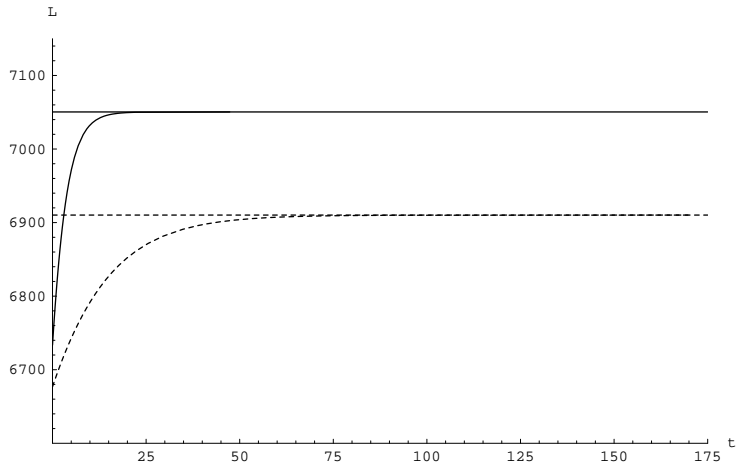}\\
  \caption{As in Fig.~\ref{lowerVc2}, but for $\Omega/2 \pi = 0.4$ and $V^2=0.1878$ much higher than the critical value $V_{c2}$.
  The times necessary to reach the steady state differ in a $500\%$ whereas the steady state values differ only in a
  $12.5\%$.
 The values of $L_0$ are the unsteady solutions $L_2$ and  $L^A_2$ at
  the same $\Omega$ and $V^2=0.1626$.}
  \label{aboveVc222}
\end{figure}

Furthermore, if we plot the non-stationary solutions for a value of
$V^2$ much higher than $V_{c2}^2$, we note that the ratio between
the temporal scales is yet bigger than the factor 3. In fact, by
setting the same value of $\Omega$ and taking $V^2=0.1878$, the
corresponding values of the stationary solutions become:
\[L_1=L_1^A= 5132,
\hspace{0.5cm} L_2= 6910, \hspace{0.5cm} L_2^A=7050,
\]
and the graphics of the solutions (\ref{SolNonStAlt}) and
(\ref{SolNonSt}) are shown in Fig.~\ref{aboveVc222}. As initial
data, we have chosen $L_0=L_{2_{|V^2=0.1626}}$ and
$L_0=L_{2_{|V^2=0.1626}}^A$ for the Vinen's equation and alternative
Vinen's one, respectively. Looking at these unsteady solutions, we
note that the solution of the alternative Vinen's equation
approaches to $L_2^A$ in a much shorter time than the other solution
requires to approach $L_2$, by a ratio of about 1:5.

Note that the time scales in Fig.~\ref{aboveVc212} [100--300
seconds] are much longer than those in Fig.~\ref{aboveVc222}
[15--75 seconds]. This is not surprising because
Fig.~\ref{aboveVc212} corresponds to a situation which is much
closer to the critical velocity $V_{c2}$ than that corresponding
to Fig.~\ref{aboveVc222}. Indeed, it is known that the dynamics
near critical points and phase transitions is much slower than in
situations far from them.

\begin{figure}
   \includegraphics[width=10cm]{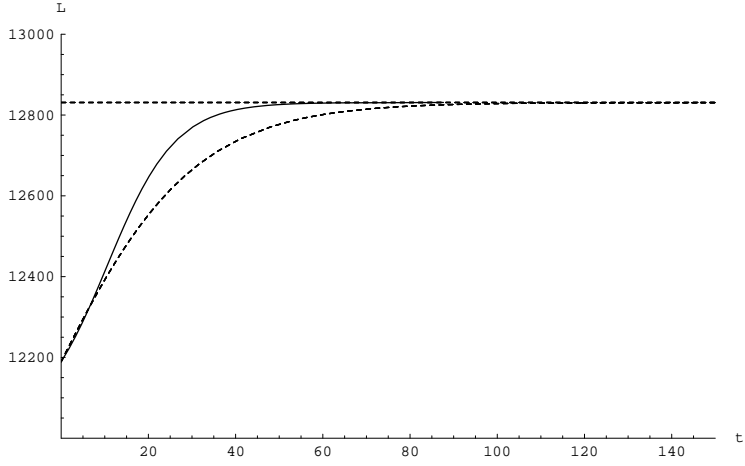}\\
  \caption{As in Fig.~\ref{lowerVc2}, for $\Omega/2 \pi = 1$ and $V^2 = 0.024<V_{c2}^2$, lower than
  the critical value $V_{c2}$. Here, the initial data $L_0$ is the value
  of $L_1$ at $\Omega/2 \pi = 0.95$ and at the same $V$.}
  \label{lowerVc2-Vcost}
\end{figure}

 \textbf{For $V$ fixed}. In the three situation
considered before, the angular velocity is always constant
($\Omega/2 \pi = 0.4$) whereas the counterflow velocity increase
from an initial value $V_0$ to a final one $V$. The global behavior
is the same when we suppose the opposite situation, that is the
counterflow velocity is maintained constant and the angular velocity
increases from an initial value $\Omega_0$ to a final one $\Omega$.
In fact, by choosing the counterflow velocity $V^2=0.024<V_{c2}^2$
and increasing $\Omega$ from $\Omega_0/2 \pi = 0.95$ to $\Omega/2
\pi = 1$ the following values for the stationary solutions are
obtained:
\[
L_1=L_1^A=12831 , \hspace{0.5cm} L_2= 11345, \hspace{0.5cm}
L_2^A=12023.
\]
For $L_0$ the value of $L_1$ at $\Omega_0/2 \pi = 0.95$ is chosen.
The plots of the unsteady solutions with the previous values is
shown in Fig.~\ref{lowerVc2-Vcost}, from which the same conclusion
of Fig.~\ref{lowerVc2} may be reached.

\begin{figure}
   \includegraphics[width=10cm]{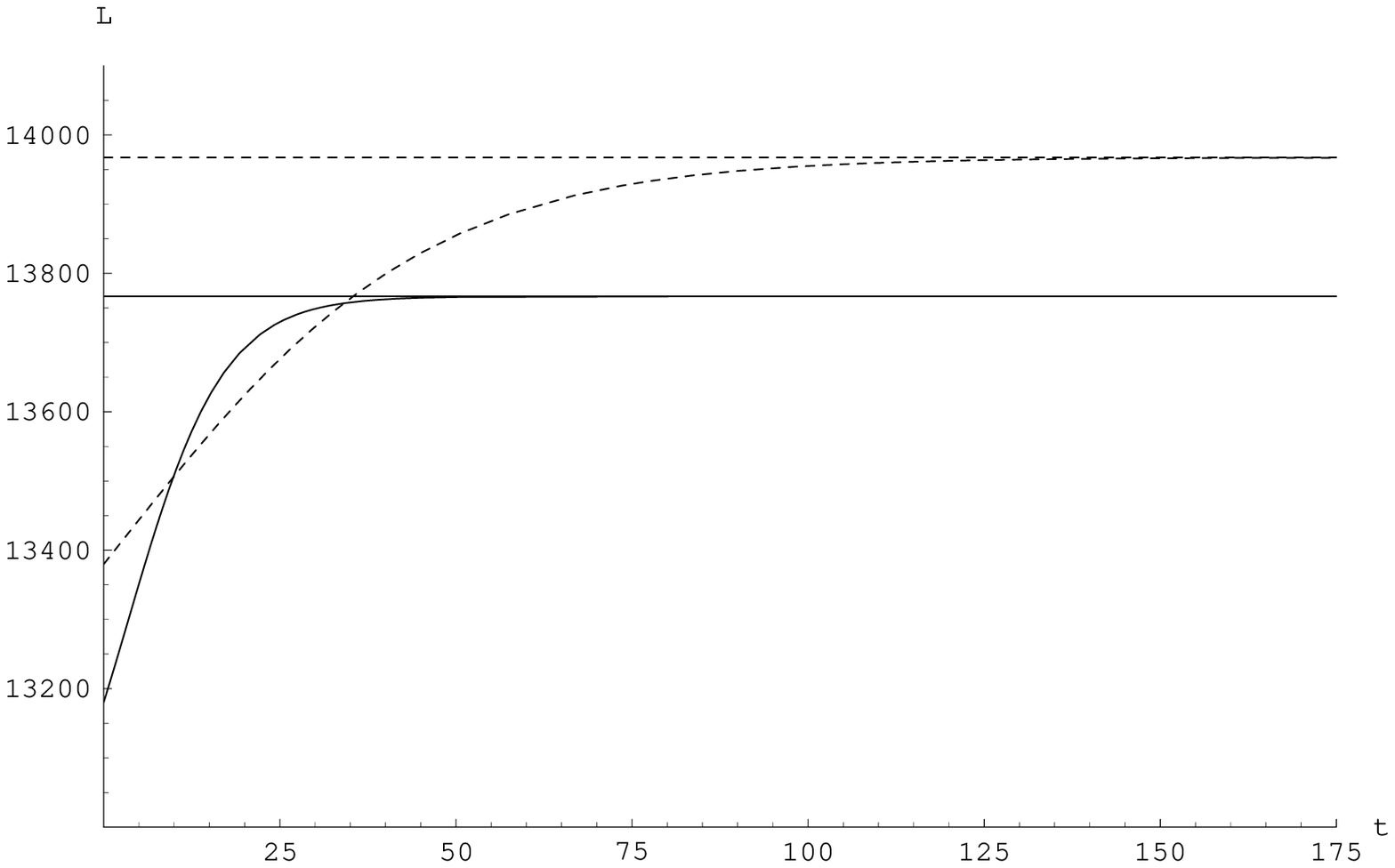}\\
  \caption{As in Fig.~\ref{lowerVc2}, for $\Omega/2 \pi = 1$ and $V^2 = 0.1626$, above the
  the critical value $V_{c2}$. The values of $L_0$ are the unsteady solutions $L_2$ and  $L^A_2$ at
  the same $V^2$ and $\Omega/2 \pi = 0.95$.}
  \label{aboveVc2-Vcost2}
\end{figure}

In Fig.~\ref{aboveVc2-Vcost2} the two non-stationary solutions
(\ref{SolNonStAlt}) and (\ref{SolNonSt}) are shown when
$V^2=0.1626>V_{c2}^2$ and $\Omega$
 increases from $\Omega_0/2 \pi = 0.95$ to $\Omega/2 \pi =1$. There, by choosing the
 initial states
$L_0=L_{2_{|\Omega/2 \pi = 0.95}}$ and  $L_0=L_{2_{|\Omega/2 \pi =
0.95}}^A$, the two solutions $L$ approach to the steady states $L_2$
and $L_2^A$:
\[
L_1=L_1^A=12831 , \hspace{0.5cm} L_2= 13967, \hspace{0.5cm}
L_2^A=13767,
\]
in different times, in agreement with the previous situations.

\section{Conclusions}
The possibility of at least two reasonable evolution equations for
the vortex line density $L$, namely (\ref{Vin}) and
(\ref{Vin-al}), was known since the early days in which Vinen
proposed them. However, detailed comparisons for them are very
scarce \cite{V3,NSF}. This was due, in part, to the fact that both
of them lead to the same form for the steady state results, namely
$L\sim V^2$, and that their unsteady solutions are not
sufficiently different to reach a definitive conclusion on their
relative merit. Here, we have carried out a detailed comparison of
an extension of both equations (\ref{Vin}) and (\ref{Vin-al}) to
the simultaneous presence of counterflow and rotation. The
extension of (\ref{Vin}) was already studied in Ref.~\cite{JM1}.
Here we have studied the analogous extension of (\ref{Vin-al}). We
have seen that in steady states the solutions of both equations,
namely (\ref{solstazL1AValue}) and (\ref{solstazL1Value}) have a
different form but in the range of available experimental results
both of them yield a satisfactory approximate description of the
experimental data. However, a deeper comparison of the
experimental errors, in (\ref{error1}) and (\ref{error2}), shows
that the description based on the alternative Vinen's equation is
slightly better than the one based on the most well-known Vinen's
equation.

A new aspect we have explored is the unsteady behavior of the
solutions of these equations. Here, both equations exhibit
remarkable differences, and we show that the solutions of the
alternative Vinen's tend much faster to their steady-state values.
In fact, this difference depends on the value of the counterflow
velocity. For $V^2=0.0626$, slightly higher than the critical
velocity $V_{c2}^2$, the time required to reach the steady state
solutions is 3 times shorter in the alternative Vinen's equation
than in the usual Vinen's equation, whereas for $V^2=0.1878$ the
difference is still more remarkable, the time scale of the
alternative Vinen's equation being 5 times shorter than that for
the usual one. Though we lack detailed experimental data on this
unsteady behavior, we know that the time required to reach the
steady state was less than 10 minutes according to Swanson {\it et
al.},  when the counterflow velocity $V$ is slightly above the
critical velocity $V_{c2}$ and it increases between two
consecutive experimental values  (see pag. 191, Ref.~\cite{SBD}).
According to the results of the Fig.~\ref{aboveVc212}, the
temporal scale of the solution of the usual Vinen's equation is
closer to the observations than the temporal scale corresponding
to the alternative equation, which tends too fast to the final
result. Thus, it seems that the usual equation is preferable on
these grounds.

    However, it must be stressed that the value of $\beta$ used in our
analysis has been $\beta \simeq 0.16$, the value corresponding to
pure counterflow, but the value of $\beta$ could depend on the
polarity of the tangle, as mentioned below expressions
(\ref{Valuealtri}). The temporal scale of the solution of the
generalized usual Vinen's equation could be set equal to the
experimental value by setting $\beta = 0.11$; instead, to adjust the
temporal behaviour of the generalized alternative Vinen's equation a
much more radical change in the value of $\beta$ should be made,
setting $\beta = 0.03$. However, such a drastic reduction of the
value of $\beta$ seems at odds with the fact that $A_1/\beta$ in the
presence of rotation is smaller than in pure counterflow, as
mentioned below (\ref{Valuealtri}). Thus, the dynamical behavior of
the usual Vinen's production terms seems more suitable than the
modified one. Our work makes also evident the need of more detailed
studies of the dependence of $\beta$ --- and other coefficients ---
on the polarization of the tangle.

\section*{Acknowledgments}
We acknowledge the support of the Acci\'{o}n Integrada
Espa\~{n}a-Italia (Grant S2800082F  HI2004-0316 of the Spanish
Ministry of Science and Technology and grant IT2253 of the Italian
MIUR). DJ acknowledges the financial support from the Direcci\'{o}n
General de Investigaci\'{o}n of the Spanish Ministry of Education
under grant BFM 2003-06033 and of the Direcci\'{o} General de
Recerca of the Generalitat of Catalonia, under grant 2005 SGR-00087.
MSM and MS acknowledge the financial support from MIUR under  grant
"PRIN 2005 17439-003" and by "Fondi 60\%" of the University of
Palermo. MS acknowledges the "Assegno di ricerca" of the University
of Palermo.

\begin{thebibliography}{99}

\bibitem{VN}
W. F. Vinen and J. J. Niemela,  J. Low Temp.~Phys.  128 (2002) 167

\bibitem{BDV}
C. F. Barenghi, R. J. Donnelly and W. F. Vinen,   Quantized Vortex
Dynamics and  Superfluid  Turbulence, Springer-Verlag  Berlin,
2001

\bibitem{V1}
W. F. Vinen,  Phys.~Rev.~B 61 (2000) 1410

\bibitem{B1}
C. F. Barenghi, J.~Phys.~Cond.~Matter 11  N 40 (1999) 7751

\bibitem{D1}
R. J. Donnelly,  J.~Phys. ~Cond.~Matter 11 N 40 (1999) 7783

\bibitem{V2}
W. F. Vinen,   Proc. Roy.Soc. ~London A 240 (1957) 493

\bibitem{V3}
W. F. Vinen,   Proc.~Roy.~Soc. London A 243 (1958) 400

\bibitem{NF}
S. K. Nemirovskii and W. Fiszdon, Rev. Mod. Phys. 67  N.1 (1995)
37

\bibitem{NSF}
S. K. Nemirovskii, G. Stamm and W. Fiszdon,  Phys. Rev. B 48
(1993) 7338

\bibitem{Schwartz}
K. W. Schwartz,  Phys. Rev. B 18 (1978) 245

\bibitem{Li}
T. Lipniacki,  in  Quantized Vortex Dynamics and  Superfluid
Turbulence, C. F. Barenghi, R. J. Donnelly and  W. F. Vinen (eds)
2001, pp. 177

\bibitem{LL}
L. D. Landau and E. M. Lifshitz,  Statistical Physics, Pergamon
Oxford, 1980

\bibitem{JLM}
D. Jou D, G. Lebon and M. S. Mongiov\`{\i}, Phys. Rev. B 66 (2002)
224509

\bibitem{KBS}
D. Kivodites, C. F. Barenghi and D. C. Samuels,  Phys. Rev. Lett.
87 (2001) 155301

\bibitem{JM-libro}
M. S. Mongiov\`{\i} and D. Jou,  in  Condensed Matter: New
Research, M. P. Das ed., Nova Science Publishers New York, 2006

\bibitem{F1}
A. P. Finne and al.,  Letters to Nature 424 (2003) 1022

\bibitem{TAB}
M. Tsubota, T. Araki and C. F. Barenghi, Phys.~Rev.~Lett. 90
(2003) 205301

\bibitem{TBAM}
M. Tsubota, C. F. Barenghi, T. Araki and A. Mitani, Phys. Rev. B
69 (2004) 134515

\bibitem{JM1}
 D. Jou and M. S. Mongiov\`{\i}, Phys. Rev. B 69 (2004) 094513

\bibitem{JM2}
D. Jou and M. S. Mongiov\`{\i}, Phys. Rev. B 72 (2005) 144517

\bibitem{F2}
R. P. Feynman, Chapter 2 in  Progress  in  Low Temperature Physics
I, North-Holland Publishing Co., Gorter C J ed. Amsterdam, 1995

\bibitem{SBD}
 C. E. Swanson, C. F. Barenghi and R. J. Donnelly, Phys.~Rev.~Lett. 50 (1983)
 190

\bibitem{YG}
E. J. Yarmchuck and W. I. Glaberson,  Phys. Rev. Lett. 41 (1978)
564

\bibitem{CCD}
D. Cheng, M. W. Cromar and R. J. Donnelly,  Phys. Rev. Lett. 31
(1973) 433

\bibitem{OG}
R. M. Ostermeir and W. I. Glaberson, J. Low Temp. Phys. 21 (1975)
191

\bibitem{JM3}
M. S. Mongiov\`{\i} and D. Jou, Phys.~Rev.~B 72 2005) 104515

\bibitem{Vinen}
W.F. Vinen,  Phys. Rev. B 61 (2000) 1410

\bibitem{JM4}
D. Jou and M. S. Mongiov\`{\i},  Phys.~Rev.~B 74 (2006) 054509

\bibitem{JM-PLA359}
 D. Jou and M. S. Mongiov\`{\i} Phys.~Lett.~A 359 (2006) 183

\bibitem{JM5}
M. S. Mongiov\`{\i} and D. Jou,  J. Phys: Cond. Matt. 17 (2005)
4423

\bibitem{SR}
K. W. Schwarz and J. R. Rozen,  Phys Rev. B 44 (1991) 7563

\bibitem{HV}
H. E. Hall and W. F. Vinen,  Proc.~Roy.~Soc. London A 238 (1956)
215

\bibitem{D2}
 R. J. Donnelly, Quantized Vortices  in
Helium II, Cambridge University Press Cambridge, 1991
\end{thebibliography}
\end{document}